\documentclass[useAMS,usenatbib]{mn2e}
\usepackage{multirow,graphicx,amsmath,amssymb,wrapfig,aas_macros}

\usepackage[usenames,dvipsnames]{color}
\usepackage{hyperref}

\title[Interacting DM: structure of MW satellites]{Dark matter--radiation interactions: the structure of Milky Way satellite galaxies}
\author[J. A. Schewtschenko, C. M. Baugh, R. J. Wilkinson, C. B\oe hm, S. Pascoli, T. Sawala]{J. A. Schewtschenko,$^{1,2}$\thanks{E-mail:
j.a.schewtschenko@dur.ac.uk}
C. M. Baugh,$^{1}$
R. J. Wilkinson,$^{2}$
C. B\oe hm,$^{2,3}$
S. Pascoli,$^{2}$
\newauthor
T. Sawala$^{4}$
\vspace*{6pt}\\
$^1$Institute for Computational Cosmology, Durham University, Durham DH1 3LE, UK\\
$^2$Institute for Particle Physics Phenomenology, Durham University, Durham DH1 3LE, UK\\
$^3$LAPTH, U. de Savoie, CNRS,  BP 110, 74941 Annecy-Le-Vieux, France\\
$^4$Department of Physics, University of Helsinki, Gustaf H\"allstr\"omin katu 2a, FI-00014 Helsinki, Finland}

\begin{document}

\date{\today}
\pubyear{2015}

\label{firstpage}
\maketitle

\begin{abstract}
In the thermal dark matter (DM) paradigm, primordial interactions between DM and Standard Model particles are responsible for the observed DM relic density. In~\cite{boehm:2014MNRAS}, we showed that weak-strength interactions between DM and radiation (photons or neutrinos) can erase small-scale  density fluctuations, leading to a suppression of the matter power spectrum compared to the collisionless cold DM (CDM) model. This results in fewer DM subhaloes within Milky Way-like DM haloes, implying a reduction in the abundance of satellite galaxies. Here we use very high resolution $N$-body simulations to measure the dynamics of these subhaloes. We find that when interactions are included, the largest subhaloes are less concentrated than their counterparts in the collisionless CDM model and have rotation curves that match observational data, providing a new solution to the ``too big to fail'' problem.
\end{abstract}


\section{Introduction}
\label{sec:intro}

The cold dark matter (CDM) model has been remarkably successful at explaining measurements of the cosmic microwave background radiation and the large-scale structure of the Universe. However, in its simplest form, the model faces challenges on small scales; the most pressing of which are the ``missing satellite''~(\citealt{moore_dark_1999,Klypin:1999uc}) and ``too big to fail''~(\citealt{BoylanKolchin:2011de}) problems. These discrepancies may indicate the need to consider a richer physics phenomenology in the dark sector, although they were first stated without the inclusion of baryonic physics.

The ``missing satellite'' problem refers to the overabundance of DM subhaloes in Milky Way (MW)-like DM haloes, compared to the observed number of MW satellite galaxies. This comparison between theory and observation requires a connection to be made between subhaloes and galaxies; in the absence of a good model for galaxy formation, this is most readily done using the halo circular velocity. Subsequent simulations that have taken into account baryonic physics suggest that a reduction in the efficiency of galaxy formation in low-mass DM haloes results in many of the excess subhaloes containing either no galaxy at all or a galaxy that is too faint to be observed (\citealt{Benson:2002, Somerville:2002,Sawala:2014,Sawala:2015}).

As the resolution of $N$-body simulations continued to improve, the ``too big to fail'' problem emerged (\citealt{BoylanKolchin:2011de}). This concerns the largest subhaloes, which should be sufficiently massive that their ability to form a galaxy is not hampered by heating of the intergalactic medium by photo-ionising photons or heating of the interstellar medium by supernovae. Simulations of vanilla CDM showed that the largest subhaloes are more massive and denser than is inferred from measurements of the MW satellite rotation curves.

The severity of the small-scale problems can be reduced if one considers the mass of the MW, which impacts the selection of MW-like haloes in the simulations but remains difficult to determine~(\citealt{Wang:2012, Cautun:2014dda, Piffl:2014, Wang:2015}). A range of alternatives to vanilla CDM have also been proposed e.g. warm DM~(\citealt{schaeffer_silk}), interacting DM~(\citealt{Boehm:2000gq,Boehm:2004th,CyrRacine:2012fz,Chu:2014lja}), self-interacting DM~(\citealt{Spergel:1999mh,Rocha:2012jg,Vogelsberger:2014pda,Buckley:2014PhRvD}), decaying DM (\citealt{Wang:2014ina}) and late-forming DM (\citealt{Agarwal:2015}). These ``beyond CDM'' models generally exhibit a cut-off in the linear matter power spectrum at small scales (high wavenumbers) that translates into a reduced number of low-mass DM haloes compared to collisionless CDM at late times.

Most numerical efforts so far to check whether such models could solve the small-scale problems have focussed on either warm DM or self-interacting DM. However, some works have studied the impact of DM scattering elastically with Standard Model particles in the early Universe; for example, with photons ($\boldsymbol{\gamma}$\textbf{CDM})~(\citealt{Boehm:2000gq,boehm_interacting_2001,Sigurdson:2004zp,Boehm:2004th,Dolgov:2013una,Wilkinson:2013kia}), neutrinos ($\boldsymbol{\nu}$\textbf{CDM})~(\citealt{Boehm:2000gq,boehm_interacting_2001,Boehm:2004th,Mangano:2006mp,Serra:2009uu,Wilkinson:2014ksa,Escudero:2015yka}) and baryons~(\citealt{Chen:2002yh,Dvorkin:2013cea,Aviles:2011ak}).

Such elastic scattering processes are intimately related to the DM annihilation mechanism in the early Universe and are thus directly connected to the DM relic abundance in scenarios where DM is a thermal weakly-interacting massive particle (WIMP). Therefore, rather than being viewed as exotica, interactions between DM and Standard Model particles should be considered as a more realistic realisation of the CDM model. Indeed, instead of assuming that CDM has no interactions beyond gravity, one can actually test this assumption by determining their impact on the linear matter power spectrum and ruling out values of the cross section that are in contradiction with observations. However, it should be noted that the strength of the scattering and annihilation cross sections can differ by several orders of magnitude, depending on the particle physics model.

The $\gamma$CDM and $\nu$CDM scenarios are characterised by the collisional damping of primordial fluctuations, which can lead to a suppression of small-scale power at late times. The collisional damping scale is determined by a single model-independent parameter: the ratio of the scattering cross section to the DM mass. The larger the ratio, the larger the suppression of the matter power spectrum. For simplicity, we assume that the scattering cross section is constant (i.e. temperature-independent), bearing in mind that temperature-dependence would give rise to the same effect but with a different value of the cross section today~(\citealt{Wilkinson:2013kia,Wilkinson:2014ksa}). In~\cite{boehm:2014MNRAS}, we confirmed that such models can provide an alternative solution to the missing satellite problem in the MW. Here we show that interacting DM could also solve the too big to fail problem\footnote{Recently, it was also demonstrated that one can simultaneously alleviate the small-scale problems of CDM by including interactions between DM and dark radiation on the linear matter power spectrum and DM self-interactions during non-linear structure formation~(\citealt{Cyr-Racine:2015ihg,Vogelsberger:2015gpr}).}.

The paper is organised as follows. In Section~\ref{sec:idmssp:sim}, we describe the setup of the $N$-body simulations that we use to study small structures. In Section~\ref{sec:tbtf}, we investigate whether interacting DM can alleviate the too big to fail problem, using MW observations. Finally, we give our conclusions in Section~\ref{sec:conc}.

\section{Simulations} \label{sec:idmssp:sim}

\begin{figure*}
\includegraphics[width=0.9\textwidth]{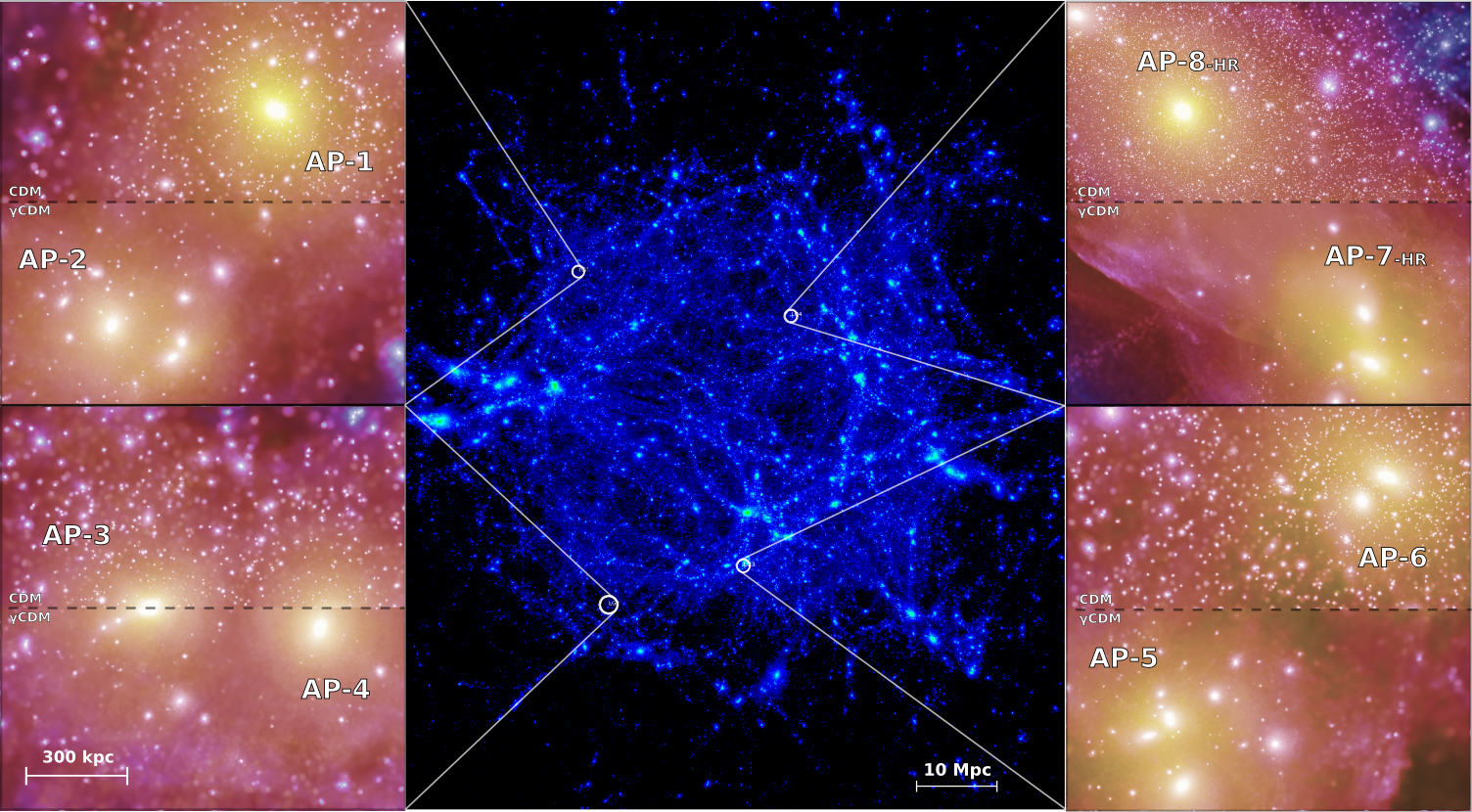}
\vspace{2ex}
\caption{The centre panel shows a projection of the DM distribution in the full $(100~ \mathrm{Mpc})^3$ \texttt{DOVE} simulation box, where the circles denote the four regions (with radii 1 $h^{-1}~{\rm{Mpc}}$) that are used for the ``zoom'' resimulations. To the left and right, each of the four Local Group candidates is rendered with the projected density encoded as brightness, where the colour scheme represents the local velocity dispersion from low (violet) to high (yellow/white).  Each of these four panels is split in half with the upper and lower halves corresponding to CDM and $\gamma$CDM with $\sigma_{\mathrm{DM}-\gamma} = 2\times 10^{-9}~\sigma_{\rm Th}~(m_{\rm DM}/{\rm GeV})$ respectively. The MW-like host haloes are labelled with the identifiers listed in Tab.~\ref{tab:idmssp:LGs}.}
\label{fig:idmssp:apostles}
\end{figure*}

While the CDM matter power spectrum predicts the existence of structures at all scales (down to earth mass haloes~(\citealt{Diemand:2005Nature,Springel:2008cc,Angulo:2009hf})), interacting DM models predict a suppression of power below a characteristic damping scale that is determined by the ratio of the DM interaction cross section to the DM mass~(\citealt{boehm_interacting_2001}). For allowed $\gamma$CDM and $\nu$CDM models~(\citealt{boehm:2014MNRAS}), the suppression occurs for haloes with masses below $10^8-10^9~M_{\odot}$. Therefore, to study the distribution and properties of structures beyond the linear regime, it is essential to carry out high-resolution $N$-body simulations.

To reach the resolution required to model the dynamics of DM subhaloes within MW-mass DM haloes, we first 
identify Local Group (LG) candidates in an $N$-body simulation of a large cosmological volume, and then resimulate the region containing these haloes at much higher mass resolution in a ``zoom'' resimulation. We use the \texttt{DOVE} cosmological simulation to identify haloes for resimulation (the criteria used to select the haloes are listed below) \citep{Sawala:2014arXiv}. The \texttt{DOVE} simulation follows the hierarchical clustering of the mass within a periodic cube of side length $100~$Mpc, using particles of mass $8.8 \times 10^{6}~M_{\odot}$ and assuming a WMAP7 cosmology.  

Following the \texttt{APOSTLE} project~\citep{Fattahi:2015arXiv}, which also uses the \texttt{DOVE} CDM simulation to identify LG candidates for study at higher resolution, we impose the following three criteria to select candidates for resimulation:
\begin{enumerate}
 \item {\bf Mass:} there should be a pair of MW and Andromeda mass host haloes, with masses  in the range $(0.5 - 2.5) \times 10^{12}~{M}_\odot$. \\

 \item {\bf Environment:} there should be no other large structures nearby, i.e. an environment with an unperturbed Hubble flow out to 4 Mpc. \\

 \item {\bf Dynamics:} the separation between the two haloes should be $800 \pm 200$ kpc, with relative radial and tangential velocities below $250$ km $\mathrm{s}^{-1}$ and $100$ km $\mathrm{s}^{-1}$ respectively.
\end{enumerate}
These criteria are more restrictive than those employed in our earlier work on the structure of haloes (\citealt{Schewtschenko:2015MNRAS}) as they also take into account the internal kinematics of the LG. We obtain four LG candidates and therefore, eight MW-like haloes. If we assume that the gravitational interaction between the LG haloes is limited, the mass, environment and dynamics\footnote{The formation process of structures is slightly delayed by the presence of DM interactions. Therefore, both the separation and the relative velocities may actually lie outside the bound set by the ``Dynamics'' criterion as the haloes are at a different point in their orbit around each other for $\gamma$CDM. However, as long as this delay between CDM and $\gamma$CDM is not too large, we essentially have the same dynamical system in both cases and the substructures within the host haloes will be unaffected.} of the haloes would not be significantly different  if we had run a  $\gamma$CDM or $\nu$CDM  version of the \texttt{DOVE} simulation.  

\begin{table}
\begin{center}
\begin{tabular}{c|ccc|l}
\hline \hline
\multirow{2}{*}{ID} & $M_{\rm vir}$ & $V_{\max}$ & $\sigma_{\mathrm{DM}-\gamma}$ \\
& $[10^{12}~M_{\odot}]$ & [$\mathrm{km}$ $\mathrm{s}^{-1}$] & $[\sigma_{\rm Th}~(m_{\rm DM}/{\rm GeV})]$ \\
 \hline \hline
 AP-1 & 1.916 & 200.3 & \multirow{2}{*}{$0,~2\times 10^{-9}$} \\
 AP-2 & 1.273 & 151.5 & \\
 \hline
 AP-3 & 0.987 & 157.9 & \multirow{2}{*}{$0,~2\times 10^{-9}$} \\
 AP-4 & 0.991 & 163.0 & \\
 \hline
 AP-5 & 2.010 & 167.5 & \multirow{2}{*}{$0,~2\times 10^{-9}$} \\
 AP-6 & 1.934 & 165.1 & \\
 \hline
 AP-7 & 1.716 & 163.7 & $0,~10^{-10},~10^{-9},$ \\
 AP-8 & 1.558 & 193.3 & $2 \times 10^{-9},~10^{-8}$ \\
\hline \hline
\end{tabular}
\end{center}
\caption{Key properties of the MW-like haloes in the zoom resimulations (Section~\ref{sec:idmssp:sim}). 
The first column specifies the \texttt{APOSTLE} identifier (ID) for each MW-like halo, while the second and third columns list the virial 
mass, $M_{\rm vir}$, and maximum circular velocity, $V_{\rm max}$, respectively (for CDM). The fourth 
column lists the different DM--photon interaction cross sections, $\sigma_{\mathrm{DM}-\gamma}$, used in 
the zoom resimulations for each LG candidate, where $\sigma_{\rm Th}$ is the Thomson cross section 
and $\sigma_{\mathrm{DM}-\gamma} = 0$ corresponds to CDM.}
\label{tab:idmssp:LGs}
\end{table}

We perform resimulations with the \texttt{P-Gadget3} $N$-body simulation code~\citep{gadget2} assuming the $\gamma$CDM model, bearing in mind that the results for $\nu$CDM would be very similar~(see~\citealt{Schewtschenko:2015MNRAS}). We use the same cosmology (WMAP7)\footnote{The fact that we are using the older WMAP7 cosmology instead of the most recent data is not a concern since we are only interested in the effects of DM interactions on a selected local environment.}, random phases and second-order LPT method~\citep{Jenkins:2010MNRAS} as~\cite{Sawala:2014arXiv}. We resimulate the four LG candidates with a particle mass  $m_{\rm part}=7.2\times10^5~M_\odot$ and a comoving softening length  $l_{\rm soft}=216$~pc. This corresponds to a mass resolution that is intermediate between levels 4 and 5 in the Aquarius simulations of \cite{Springel:2008cc} (level 1 being the highest resolution). We also resimulate the two host haloes in one of our LG Candidates (AP-7/AP-8) at an even higher resolution ($m_{\rm part}=6\times10^4~M_\odot$, $l_{\rm soft}=94$~pc; which is comparable to Aquarius level 3). These simulations (denoted with the suffix \texttt{-HR}) are used to confirm that our results have converged\footnote{While the properties of some small subhaloes may vary from one resolution level to another (due to minor changes in their formation and accretion histories), the overall scatter for all subhaloes remains the same and thus can be considered to have converged.} and allow us to obtain more reliable predictions for the innermost parts of the halo. Substructures within the host haloes are located using the \texttt{AMIGA} halo finder~\citep{ahf_refs}.

We run resimulations for zero interaction cross section, which corresponds to collisionless CDM, and for a selection of DM--photon interaction cross sections, as listed in Tab.~\ref{tab:idmssp:LGs}. We note that the DM--photon interaction cross section value of $\sigma_{\mathrm{DM}-\gamma} = 2\times 10^{-9}~\sigma_{\rm Th}~(m_{\rm DM}/{\rm GeV})$ was shown to solve the missing satellite problem in \cite{boehm:2014MNRAS}, in the absence of baryonic physics effects.

Fig.~\ref{fig:idmssp:apostles} shows the projected matter density in the \texttt{DOVE} simulation box~\citep{Sawala:2014arXiv} (central panel) along with renderings of the four LG candidates from the resimulations. The eight MW-like haloes are listed in Tab.~\ref{tab:idmssp:LGs} with their respective properties for CDM. Halo properties for $\gamma$CDM vary only slightly (within a few percent) from those in CDM for the cross sections considered here.

\section{Results}
\label{sec:tbtf}

\begin{figure}
\begin{center}
\includegraphics[width=.5\textwidth,clip,trim=4ex 0ex 4ex 10ex]{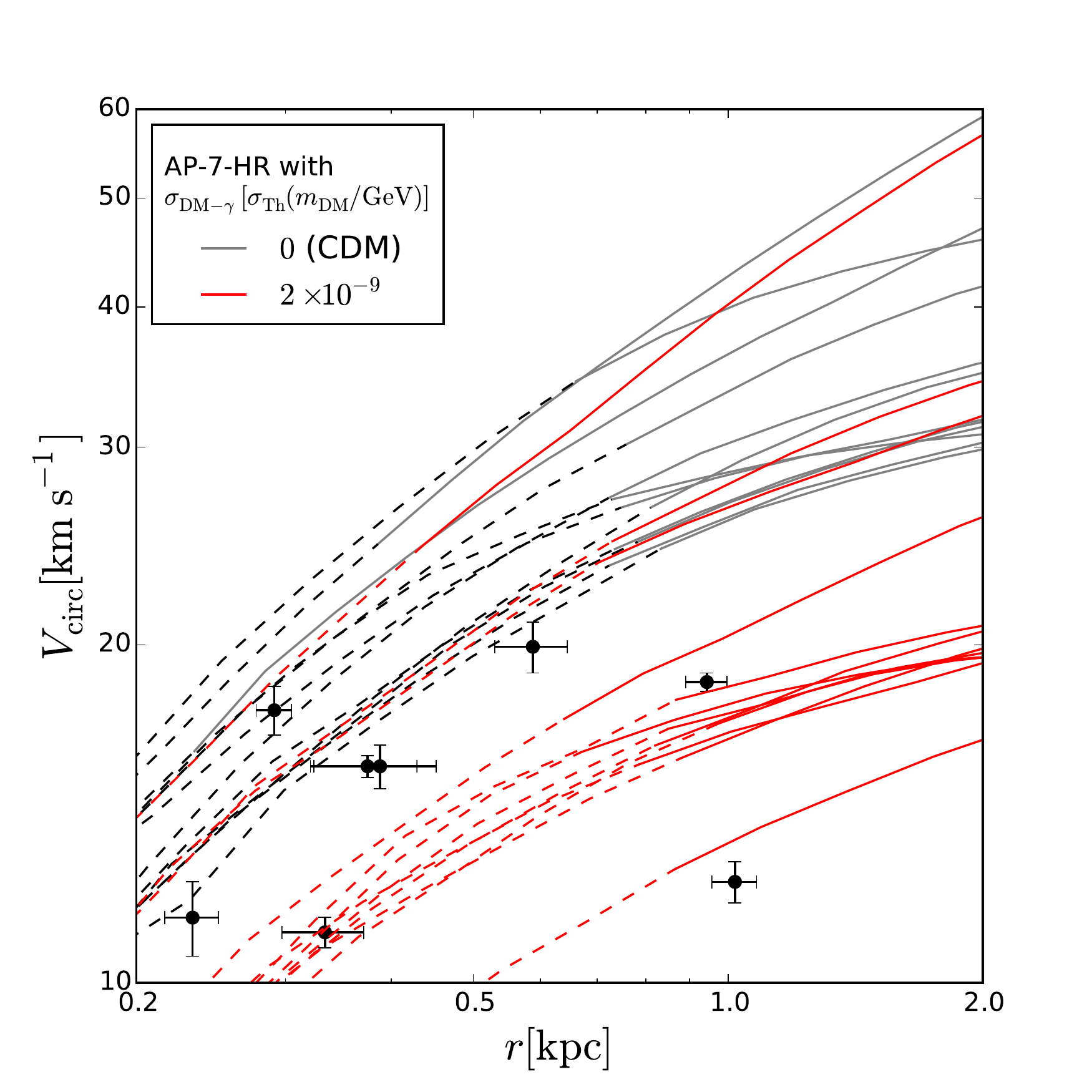}
\includegraphics[width=.5\textwidth,clip,trim=4ex 0ex 4ex 0ex]{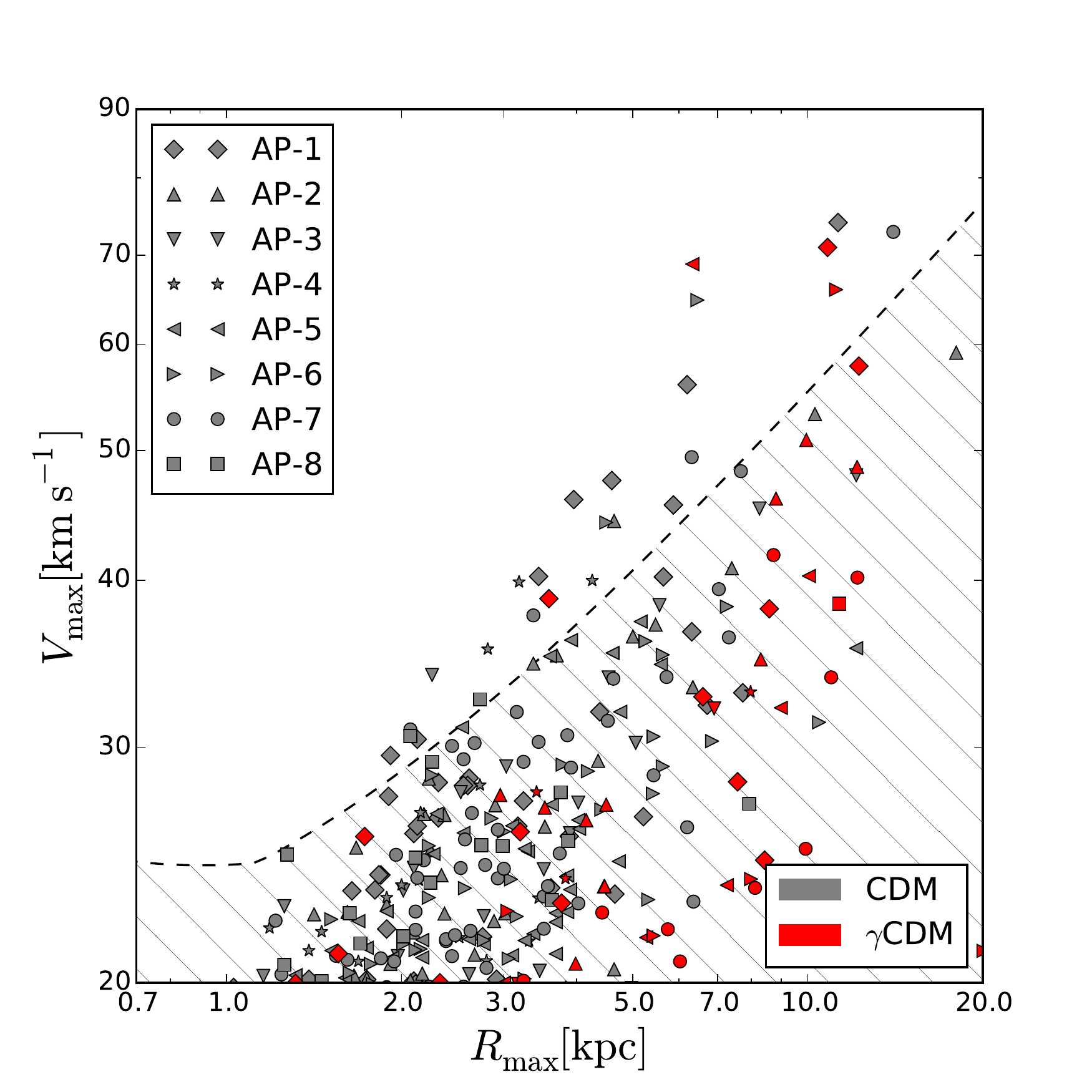}
\caption{Top: the circular velocity, $V_{\rm circ}$, versus radius, $r$, for the eleven most massive subhaloes in AP-7-HR for CDM (grey lines) and for $\gamma$CDM with $\sigma_{\mathrm{DM}-\gamma} = 2 \times 10^{-9}~\sigma_{\rm Th}~(m_{\rm DM}/{\rm GeV})$ (red lines). The dashed lines indicate where $V_{\rm circ}$ can still be measured from the simulation but convergence cannot be guaranteed, according to the criteria set out by~\citet{power:2003MNRAS}. The data points correspond to the observed MW satellites with 1$\sigma$ error bars~\citep{Wolf:2010MNRAS}. Bottom: the $V_{\rm max}$ versus $R_{\rm max}$ results for all eight MW-like haloes, with the same scattering cross sections as in the top panel. The hatched region marks the 2$\sigma$ confidence interval for the observed MW satellites. $V_{\rm max}$ is derived from the observed stellar line-of-sight velocity dispersion, $\sigma_{\star}$, using the assumption that $V_{\rm max} = \sqrt{3} \sigma_{\star}$~\citep{Klypin:1999uc}.}
\label{fig:idmssp:tbtf_gcdm}
\end{center}
\end{figure}

We now explore the too big to fail problem and show how the theoretical predictions and observations can be reconciled  by including DM interactions beyond gravity. 

The too big to fail problem is illustrated in the top panel of Fig.~\ref{fig:idmssp:tbtf_gcdm}. Here the rotation curves of the 11 most massive subhaloes\footnote{We have included three more simulated subhaloes than the observed number of dwarf satellites as the most massive subhaloes are considered statistical outliers like the Magellanic clouds, which have been omitted in this study.} in the CDM resimulation of the halo AP-7-HR clearly lie above the measurements for the ``classical'' dwarf spheroidal satellites in the MW taken from~\cite{Wolf:2010MNRAS}. In general, one can see that the largest subhaloes in CDM simulations have a higher circular velocity, $V_{\rm circ}$, and therefore more enclosed (dark) matter, than is observed for a given radius.

In the case of $\gamma$CDM, the rotation curves of the most massive satellites are shifted to lower circular velocities, indicating that there is less (dark) matter enclosed within a given radius. Alternatively, one can interpret this as a lower central density or concentration for the haloes in $\gamma$CDM, as seen in \cite{Schewtschenko:2015MNRAS}. 

The circular velocity profiles shown in the top panel of Fig.~\ref{fig:idmssp:tbtf_gcdm} are plotted using different line styles. The transition occurs at the scale determined by the convergence criteria devised by \cite{power:2003MNRAS}. At smaller radii (dashed lines), the velocity profiles are not guaranteed to have converged. However, the key point here is that the CDM and $\gamma$CDM resimulations have the same resolution and yet show a clear difference at all radii plotted, with a shift to lower circular velocities for the haloes in $\gamma$CDM.

The bottom panel of Fig.~\ref{fig:idmssp:tbtf_gcdm} presents a related view of the too big to fail problem; this time showing the peak velocity in the rotation curve, $V_{\rm max}$, and the radius at which this occurs, $R_{\rm max}$. The hatched region indicates the $2\sigma$ range inferred for the observed MW satellites, assuming that these are DM-dominated and fitting NFW profiles~(\citealt{Navarro:1997ApJ}) to the rotation curve measurements. We allow the halo concentration parameter to vary, following the same technique and assumptions as described in \cite{Klypin:1999uc}\footnote{A plot with the confidence bands for each of the MW satellites can be found in the provided online material and the augmented content of this paper.}.

Again, the collisionless CDM model predicts satellites that lie outside the $2\sigma$ range compatible with observations. Additionally, for CDM, there are many more subhaloes within the range of $V_{\rm max}$--$R_{\rm max}$ plotted than there are observed satellites. The abundance of massive, concentrated subhaloes varies depending on the mass and formation history of the host halo; however, for all the MW-like candidates studied, CDM exhibits a too big to fail problem, which is reduced in the case of $\gamma$CDM.

\begin{figure}
\begin{center}
\includegraphics[width=.5\textwidth,clip,trim=4ex 0ex 4ex 10ex]{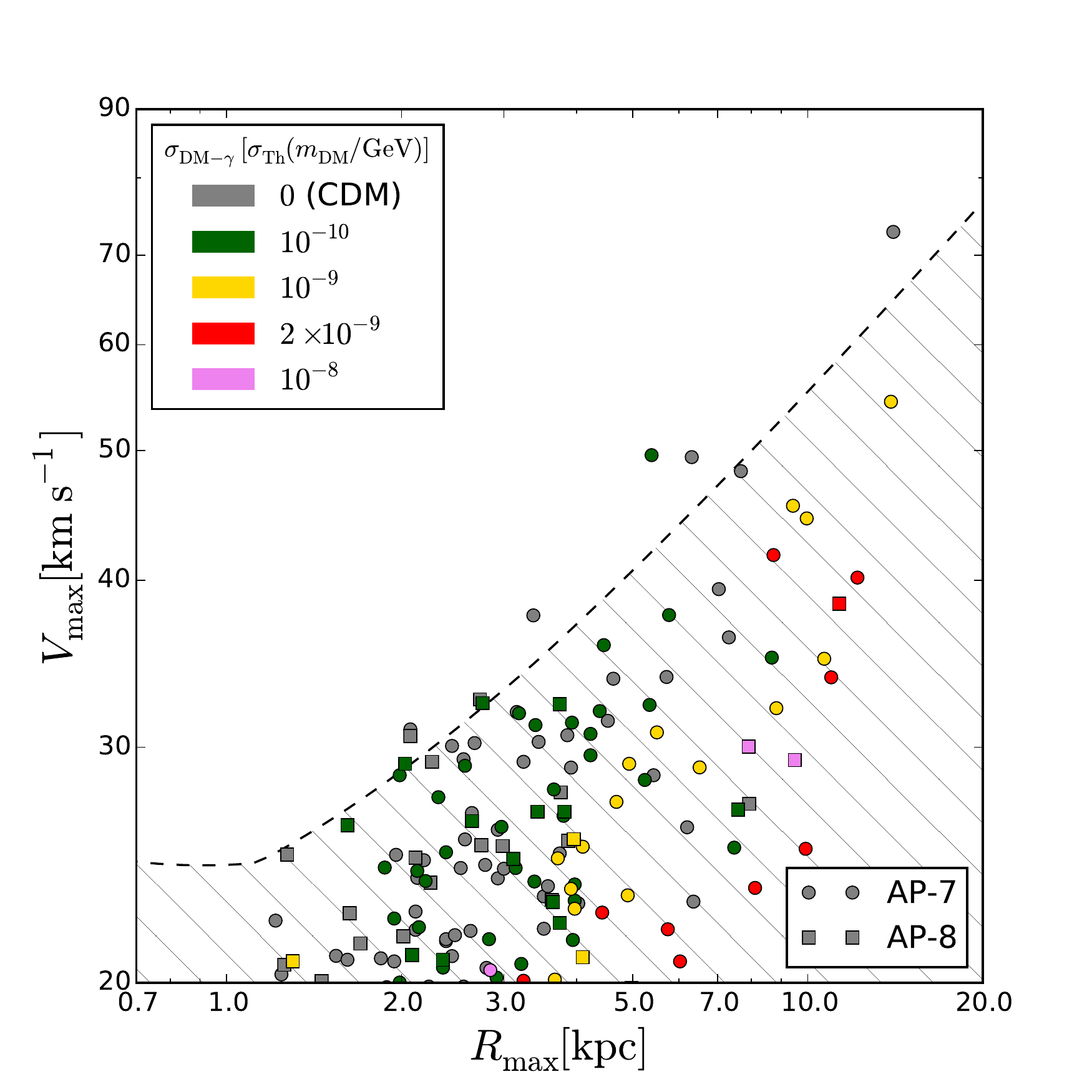}
\caption{The $V_{\rm max}$ versus $R_{\rm max}$ results for a range of DM--photon interaction cross sections using the AP-7 and AP-8 haloes. As in Fig.~\ref{fig:idmssp:tbtf_gcdm}, the hatched region marks the 2$\sigma$ confidence interval for the observed MW satellites, following the methodology of \citet{Klypin:1999uc}.}
\label{fig:idmssp:tbtf_gcdm2}
\end{center}
\end{figure}

In Fig.~\ref{fig:idmssp:tbtf_gcdm2}, we present the results for AP-7 and AP-8 to show the impact of varying the DM--photon interaction cross section. As the cross section is increased, the predicted $V_{\rm max}$ values fall and shift to larger $R_{\rm max}$. This brings the model predictions well within the region compatible with the observational results and also reduces the number of satellites with such rotation curves.  Therefore, one can clearly see that interacting DM can alleviate the too big to fail problem for a cross section $\sigma_{\mathrm{DM}-\gamma} \simeq 10^{-9}~\sigma_{\rm Th}~(m_{\rm DM}/{\rm GeV})$ that also solves the missing satellite problem~(\citealt{boehm:2014MNRAS}).

\section{Conclusion}
\label{sec:conc}

There are a multitude of solutions proposed to overcome the small-scale ``failures'' of cold dark matter (CDM); namely, the ``missing satellite'' and ``too big to fail'' problems. Within the collisionless CDM model, these explanations fall into two camps: i) invoking baryonic physics to reduce the efficiency of galaxy formation in low-mass DM haloes \citep{Sawala:2014arXiv,2015arXiv151101098S}, and ii) exploiting the uncertainty in the mass of the Milky Way (MW) DM halo \citep{Wang:2012}. Both problems can be diminished using one or both of the above.

Solutions in which the properties of the DM are varied have also been explored. \cite{Lovell:2013ola} showed that replacing CDM by a warm DM particle of mass $1.5$~keV leads to a reduced abundance of subhaloes in MW-like haloes, and massive subhaloes that are less concentrated than their CDM counterparts, matching observations of the internal dynamics of the MW satellites. \cite{Vogelsberger:2014pda} investigated the impact of self-interacting DM on the properties of satellite galaxies, finding little change in the global properties of the galaxies but variation in their structure.  

Here we have investigated the impact of interactions between DM and radiation on the structure of the MW
satellites. Such interactions are well-motivated and may have helped to set the abundance of DM inferred in 
the Universe today \citep{Boehm:2003hm,Peter:2012}; sometimes called the WIMP miracle. As well as its physical basis, this model has the attraction that it is as simple to simulate as CDM. The interactions took place in the early Universe when the densities of DM and radiation were much higher, and are negligible over the time period covered by the simulation. The DM particles are still cold, so there are no issues relating to particle velocity distributions, as would arise in high-resolution simulations of warm DM, particularly for lighter candidates. The only change compared to a CDM simulation is the modification to the matter power spectrum in linear perturbation theory; the DM--radiation interactions result in a damping of the matter power spectrum on small scales. 

We have used high resolution $N$-body simulations of DM haloes, which have passed a set of Local Group selection criteria, to show the impact of DM--radiation interactions on the structure of massive subhaloes. Increasing the interaction cross section reduces the mass enclosed within a given radius in the subhaloes, compared to their CDM counterparts, as suggested by our results for a wider population of DM haloes~(\citealt{Schewtschenko:2015MNRAS}). When combined with our earlier work showing that stronger interactions also lead to a reduction in the number of MW subhaloes~(\citealt{boehm:2014MNRAS}), we find that a model with an elastic scattering cross section of $\simeq 10^{-9}~\sigma_{\rm Th}~(m_{\rm DM}/{\rm GeV})$ can solve both of these small-scale problems of CDM. The next step will be to include baryonic physics. This will not alter the qualitative conclusions of our papers, but will relax the constraints on the DM--radiation scattering cross section.


\section*{Acknowledgements}

We thank V. Springel for providing access to the \texttt{P-Gadget3} code and the \texttt{bPic} rendering code, A. Jenkins for sharing his IC generator code for the ``zoom'' resimulations with us and J. Halley for helpful discussions. JAS is supported by a Durham University Alumnus Scholarship and RJW is supported by the STFC Quota grant ST/K501979/1. This work was supported by the STFC (grant numbers ST/F001166/1, ST/G000905/1 and ST/L00075X/1) and the European Union FP7 ITN INVISIBLES (Marie Curie Actions, PITN-GA-2011-289442). This work was additionally supported by the European Research Council under ERC Grant ``NuMass'' (FP7-IDEAS-ERC ERC-CG 617143). It made use of the DiRAC Data Centric system at Durham University, operated by the ICC on behalf of the STFC DiRAC HPC Facility (www.dirac.ac.uk). This equipment was funded by BIS National E-infrastructure capital grant ST/K00042X/1, STFC capital grant ST/H008519/1, STFC DiRAC Operations grant ST/K003267/1 and Durham University. DiRAC is part of the National E-Infrastructure. SP also thanks the Spanish MINECO (Centro de excelencia Severo Ochoa Program) under grant SEV-2012-0249. CMB acknowledges a research fellowship from the Leverhulme Trust.

\section*{Augmented content}
\begin{wrapfigure}{l}{.14\textwidth}
\includegraphics[width=.16\textwidth]{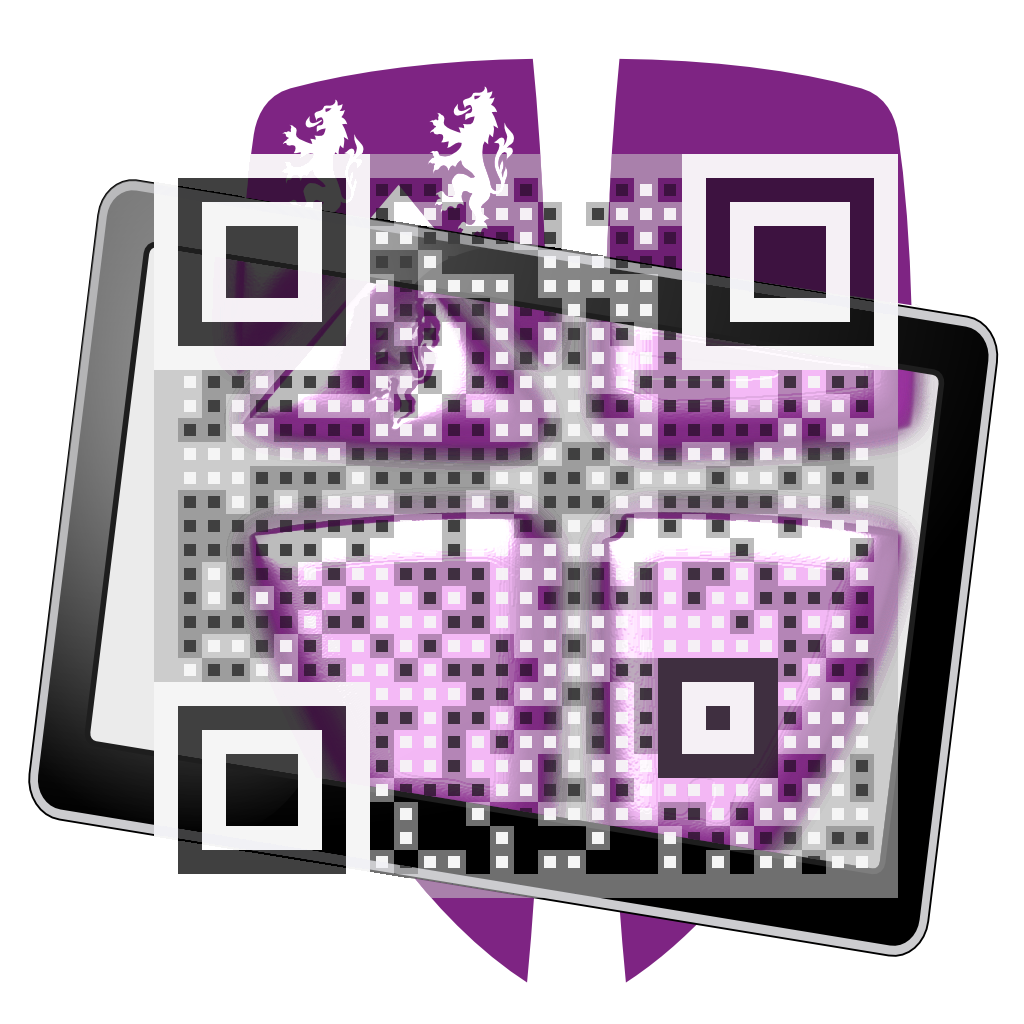}
\end{wrapfigure} 
This paper provides additional multimedia and interactive content embedded in an augmented reality using the open DARO framework for mobile devices. In order to access the data, you need to use a DARO-compatible browser (\href{http://icc.dur.ac.uk/~daro}{\texttt{http://icc.dur.ac.uk/$\sim$daro}}) and scan the DARO QR code printed here.\\

\bibliographystyle{mn2e}	
\bibliography{IDM_TBTF}

\label{lastpage}

\end{document}